\documentclass[runningheads]{llncs}
\usepackage{graphicx}
\usepackage{cite}
\usepackage{multirow}
\usepackage{xcolor}

\begin{document}

\newcommand{\modified}[1]{\textcolor{red}{#1}}

\title{Simulation-Driven Training of Vision Transformers Enabling Metal Segmentation in X-Ray Images}
%
%
\author{Fuxin Fan\inst{1} \and
Ludwig Ritschl\inst{2} \and
Marcel Beister\inst{2} \and
Ramyar Biniazan\inst{2} \and
Bj\"orn Kreher\inst{2} \and
Tristan M. Gottschalk\inst{1,2} \and
Steffen Kappler\inst{2} \and
Andreas Maier\inst{1}}

%
\institute{Pattern Recognition Lab, Friedrich-Alexander-Universit\"at Erlangen-N\"urnberg, Erlangen 91058, Germany \and
Siemens Healthcare GmbH, Forchheim, Germany
}
\maketitle              
\begin{abstract}
In several image acquisition and processing steps of X-ray radiography, knowledge of the existence of metal implants and their exact position is highly beneficial (e.g. dose regulation, image contrast adjustment). Another application which would benefit from an accurate metal segmentation is cone beam computed tomography (CBCT) which is based on 2D X-ray projections. Due to the high attenuation of metals, severe artifacts occur in the 3D X-ray acquisitions. The metal segmentation in CBCT projections usually serves as a prerequisite for metal artifact avoidance and reduction algorithms. Since the generation of high quality clinical training is a constant challenge, this study proposes to generate simulated X-ray images based on CT data sets combined with self-designed computer aided design (CAD) implants and make use of convolutional neural network (CNN) and vision transformer (ViT) for metal segmentation. Model test is performed on accurately labeled X-ray test datasets obtained from specimen scans. The CNN encoder-based network like U-Net has limited performance on cadaver test data with an average dice score below 0.30, while the metal segmentation transformer with dual decoder (MST-DD) shows high robustness and generalization on the segmentation task, with an average dice score of 0.90. Our study indicates that the CAD model-based data generation has high flexibility and could be a way to overcome the problem of shortage in clinical data sampling and labelling. Furthermore, the MST-DD approach generates a more reliable neural network in case of training on simulated data.

\end{abstract}

\section{Introduction}
Metallic implants are utilized for fixation of trauma fractures during orthopaedic surgery. However, their existence inside the body poses a severe challenge for the generation of X-ray images or acquiring fluoroscopic image sequences for cone beam computed tomography (CBCT). On the one hand the existence of metal complicates automatic dose exposure during acquisition and must be treated adequately during image processing in classical X-ray radiography. On the other hand metal has strong negative impact on image quality in CBCT. Due to the high attenuation of metallic objects, strong artifacts can appear which can limit the assessment of bone-implant integration and success of fusion procedures \cite{netto2022implant}. To reduce the impact of metal artifacts, many algorithms have been developed, such as metal artifact avoidance (MAA) and metal artifact reduction (MAR) method. MAA tends to optimize scan geometry based on limited projections to minimizes metal-induced biases in the projection data \cite{wu2020c}. MAR focuses on the removal of artifacts\cite{katsura2018current}. Both methods rely on accurate segmentation of metals \cite{wu2020c, katsura2018current, zhang2007reducing, bazalova2007correction, yu2007segmentation, yu2009metal, li2010metal, hegazy2019u, shi2021single, gottschalk2021view} which is performed either directly on projections or on reconstructed volume. Most of the commercially available MAR algorithms are based on the volume based metal segmentation \cite{katsura2018current}. It is well known that detecting metal objects is problematic in CT volumes when they are outside of the scanner's field-of-view (FOV). These limitations further motivate us to design a pure projection-based metal segmentation algorithm in this study.

With the development of deep learning approaches, neural network models are widely used in image segmentation tasks and among different model architectures, U-Net \cite{ronneberger2015u} based models have been applied mostly in metal segmentation on projection domain such as the work by Hegazy et al. \cite{hegazy2019u, gottschalk2021view}. In their study, neural network models have been trained and tested on a finite quantity of clinical and cadaver scans, which makes their algorithm highly dependent on training data and, therefore,  the generalization of their model is not guaranteed.

Traditionally training a segmentation algorithm for this task requires clinical image data consisting a combination of various types of metal implants, different body regions and different projection geometries. These data would need to undergo a high quality annotation process which might still be inaccurate in case of complex anatomical structures. In such scenarios, simulated projections have great advantages because of their accurate annotations, variety projections and the possibility to utilize different types of implants. In our case we combined the forward projection of CT data sets with the forward projection of metal volumes. Although simulation frameworks can provide a theoretical infinite number of training data, one should consider the differences between such data and real clinical images. Therefore the DeepDRR framework \cite{unberath2018deepdrr} is used by considering the physical properties of clinical data. Furthermore, vision transformer (ViT) models are well known to be able to transform between tasks \cite{dosovitskiy2020image}. Models such as SETR \cite{zheng2021rethinking}, TransUNet \cite{chen2021transunet}, CoTr \cite{xie2021cotr} have shown great potential to construct robust segmentation networks. This might be beneficial for transferring the algorithms performance from simulated X-ray projection to real projection data.

The main contributions of this work can be summarized as follows:

1) Computer-aided design (CAD) model based multi-metal projections generation. This is the first study to use CAD models to generate large numbers of projections for metallic implant segmentation.

2) Introduction of MST-DD, consisting of ViT encoder and dual connected CNN decoders. This is the first model using ViT in metal segmentation task.

\section{Materials and Method}
\subsection{Data generation}
\begin{figure}[hbt!]
\includegraphics[width=\textwidth]{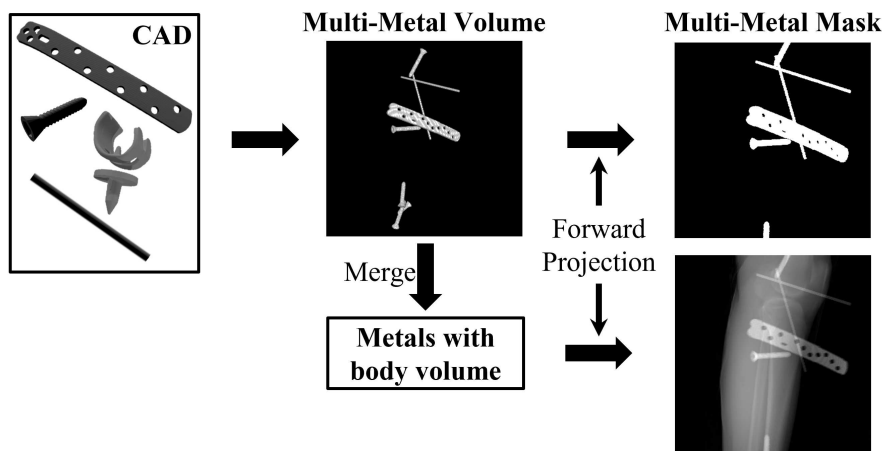}
\caption{The pipeline of multi-metal projection generation} \label{MetalMap}
\end{figure}

The pipeline for the generation of simulated projections is illustrated in Fig.~\ref{MetalMap}. The CAD models of implants, such as K-wires, screws and curved plates with holes are drawn using the software AutoCAD. Each of the models is transferred into a 3D binary image. 14 knee CT volumes have been selected from the SICAS medical image repository \cite{kistler2013virtual}. All volumes are rescaled to a voxel side length of 0.5\,mm and a volume size of 1000\,$\times$\,600\,$\times$\,600, among which 600 continues slices are randomly chosen to merge with one multi-metal volume. To generate the multi-metal volume with the size of 600\,$\times$\,600\,$\times$\,600, implants are randomly selected, rotated and translated. The Hounsfield units (HU) value for each implant is randomly chosen between 3000 and 8000. Because in reality implants typically have intersection with bone, one constraint for the translation is that the corresponding HU value at the location of the target knee volume is above 500 HU. Afterwards, the multi-metal volume is inserted into one target knee volume. In total, 50 volumes with different metal distributions and anatomical background are generated and forward projected afterwards.

Monochromatic and polyhromatic projections are simulated by CONRAD \cite{maier2013conrad} and DeepDRR framework \cite{unberath2018deepdrr}, respectively. The distance between detector and source is 1164 mm and the distance between detector and isocenter is 700 mm. The detector has the size of 976\,$\times$\,976 and each pixel is 0.305 mm\,$\times$\,0.305 mm. For one volume, 60 projections are generated with an increased rotation angle of 6$^\circ$. For the simulation of polychromatic projection, the volume is decomposed into 4 different materials depending on their HU value: air (\textless 800 HU), soft tissue ([-800, 350] HU), bone ((350, 2000] HU) and metallic implants (\textgreater 2000 HU). After Poisson noise injection and log transform, all projections are normalized. Furthermore, we use the gamma function for brightness adjustment as data augmentation. The gamma value is randomly chosen between 0.7 and 1.3. For monochromatic and polychromatic dataset, 50 sets of projections are simulated for each dataset. 45 sets among them are used for training and 5 sets are for validation. 

As test dataset we use real X-ray projection data taken from an experimental cadaver study. The cadaver scans were acquired using a mobile C-arm system with CBCT capability. All scans were acquired with and without metallic implants while keeping the scan setup in a constant position. This helps to generate highly accurate metal annotations by subtracting two corresponding scans. 10 scans are used as test data set while each scan includes 400 projections.

\subsection{Neural networks}
\begin{figure}[hbt!]
\includegraphics[width=\textwidth]{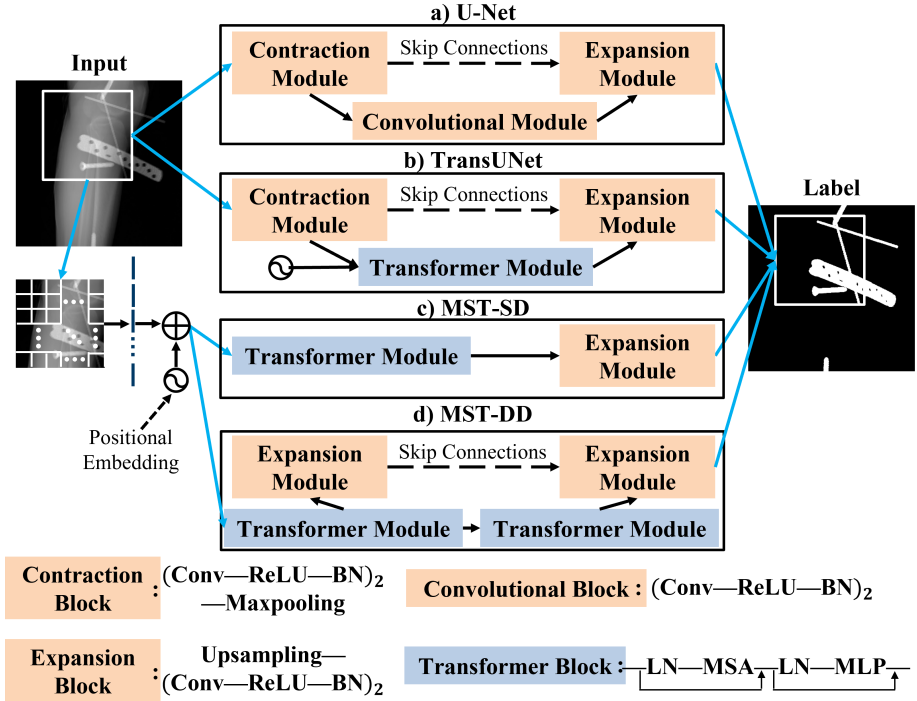}
\caption{Four Neural networks structures. a) U-Net. b) TransUNet. c) MST-SD. d) MST-DD. They are constructed by different modules which consist of several corresponding blocks.} \label{Networks}
\end{figure}
The neural networks used in this work are demonstrated in Fig.~\ref{Networks}. The basic network modules are classified into four categories: contraction module, convolutional module, expansion module and transformer module. The contraction module in this work includes four blocks and every block consists of two sequential convolutional layers, each of which is followed with a ReLU layer and a batch normalization (BN) layer. Finally, the output of these layers goes to a max pooling layer. After that the filter size is doubled for convolutional layers in next block. The convolutional module and the expansion module have one and four blocks respectively, both of which have similar structures as the contraction block, with the difference that there is no pooling layer in the convolutional block and the expansion block. Moreover, the upsampling layer is added in the beginning of the first convolutional layer of the expansion block. The filter size for the convolutional block is doubled as its previous contraction block in U-Net (Fig.~\ref{Networks}(a)), while the filter size is halved in the expansion blocks after each upsampling layer. In this work, the transformer module used in TransUNet (Fig.~\ref{Networks}(b)) and metal segmentation transformer with single decoder (MST-SD) (Fig.~\ref{Networks}(c)) has 12 transformer blocks, each of which starts with a layer normalization (LN), followed by a multi-headed self-attention layer, another LN and finally one fully connected layer (multi-layer perception). The input before each LN will be summed with the output of MSA and MLP, respectively. For MST with dual decoder (MST-DD) (Fig.~\ref{Networks}(d)), both transformer modules have 6 blocks, and they are followed by two expansion modules with skip connections. In the end, Sigmoid function is used as the active function for the output of the all networks.

In this work, the input image has the size of 512\,$\times$\,512, which is a patch randomly sampled from the original image. The further hyperparameters for the networks are as follows The filter size of the first contraction block and the last expansion block is 64. The stride size for maxpooling and upsampling is 2. The positional encoding for ViT is from 1 to the sequence size. The patch size for splitting the input image is 16\,$\times$\,16 and each patch is flattened and embedded into 256. There are four heads for the multi-headed self attention layer. The filter size is 512 for the first expansion block in TransUNet, MST-SD and MST-DD.

All the networks are trained for maximum 50 epochs. The default loss function is binary cross entropy while dice loss is incorporated for comparison. The total number of training and validation projections for one epoch is 2700 and 300, respectively. For each projection, 5 patches are sampled for training and validation. We used Adam optimizer with an initial learning rate of 0.001 and an exponential decay of 0.95 after one epoch. The evaluation metrics are dice score, precision and recall for cadaver test data.

\section{Results}
The evaluation results are summarized in Tab.~\ref{table_mono} and Tab.~\ref{table_poly}. All four networks are trained on monochromatic and polychromatic projections separately. For the models trained on monochromatic projections, both U-Net and TransUNet have poor performance with average dice score lower than 0.30 and average precision lower than 0.20. MST-SD model generates slightly better predictions, but the average dice score is still below 0.70 with a high variance of 0.16. By using dual decoder, the average dice score will increase to a value above 0.85 with a variance of 0.09 and, finally, the change of loss function to dice makes the MST-DD achieve 0.90 average dice score and a variance of 0.03.

\begin{table}[hbt!]
\caption{Evaluation results for networks trained by monochromatic projections.}\label{table_mono}
\begin{tabular}{c|lp{0.02cm}p{2.3cm}p{2.3cm}p{2.3cm}}
\hline
 Test&&& Avg. Dice & Avg. Precision & Avg. Recall\\
\hline
\multirow{5}{*}{Cadaver}&U-Net && 0.21\,$\pm$\,0.12 & 0.12\,$\pm$\,0.09 & 0.97\,$\pm$\,0.02 \\
&TranUNet && 0.28\,$\pm$\,0.13 & 0.17\,$\pm$\,0.09 & 0.93\,$\pm$\,0.04 \\
&MST-SD && 0.69\,$\pm$\,0.16 & 0.64\,$\pm$\,0.22 & 0.81\,$\pm$\,0.10 \\
&MST-DD && 0.86\,$\pm$\,0.09 & 0.95\,$\pm$\,0.01 & 0.80\,$\pm$\,0.13 \\
&MST-DD (Dice Loss) && 0.90\,$\pm$\,0.03 & 0.88\,$\pm$\,0.02 & 0.92\,$\pm$\,0.07 \\
\hline
\end{tabular}
\end{table}

For model trained on polychromatic projections, the average dice score for the U-Net and TransUNet is below 0.35. MST-SD has higher average dice score of 0.76. MST-DD models have the same dice score of 0.89, but with dice loss the variance for dice score is 0.02 lower.

\begin{table}[hbt!]
\caption{Evaluation results for networks trained by polychromatic projections.}\label{table_poly}
\begin{tabular}{c|lp{0.02cm}p{2.3cm}p{2.3cm}p{2.3cm}}
\hline
 Test&&& Avg. Dice & Avg. Precision & Avg. Recall\\
\hline
\multirow{5}{*}{Cadaver}&U-Net && 0.27\,$\pm$\,0.14 & 0.17\,$\pm$\,0.10 & 0.88\,$\pm$\,0.08 \\
&TranUNet && 0.33\,$\pm$\,0.18 & 0.22\,$\pm$\,0.15 & 0.95\,$\pm$\,0.03 \\
&MST-SD && 0.76\,$\pm$\,0.15 & 0.95\,$\pm$\,0.01 & 0.66\,$\pm$\,0.19 \\
&MST-DD && 0.89\,$\pm$\,0.07 & 0.94\,$\pm$\,0.02 & 0.85\,$\pm$\,0.11 \\
&MST-DD (Dice Loss) && 0.89\,$\pm$\,0.05 & 0.89\,$\pm$\,0.02 & 0.90\,$\pm$\,0.09 \\
\hline
\end{tabular}
\end{table}

\begin{figure}[hbt!]
\includegraphics[width=\textwidth]{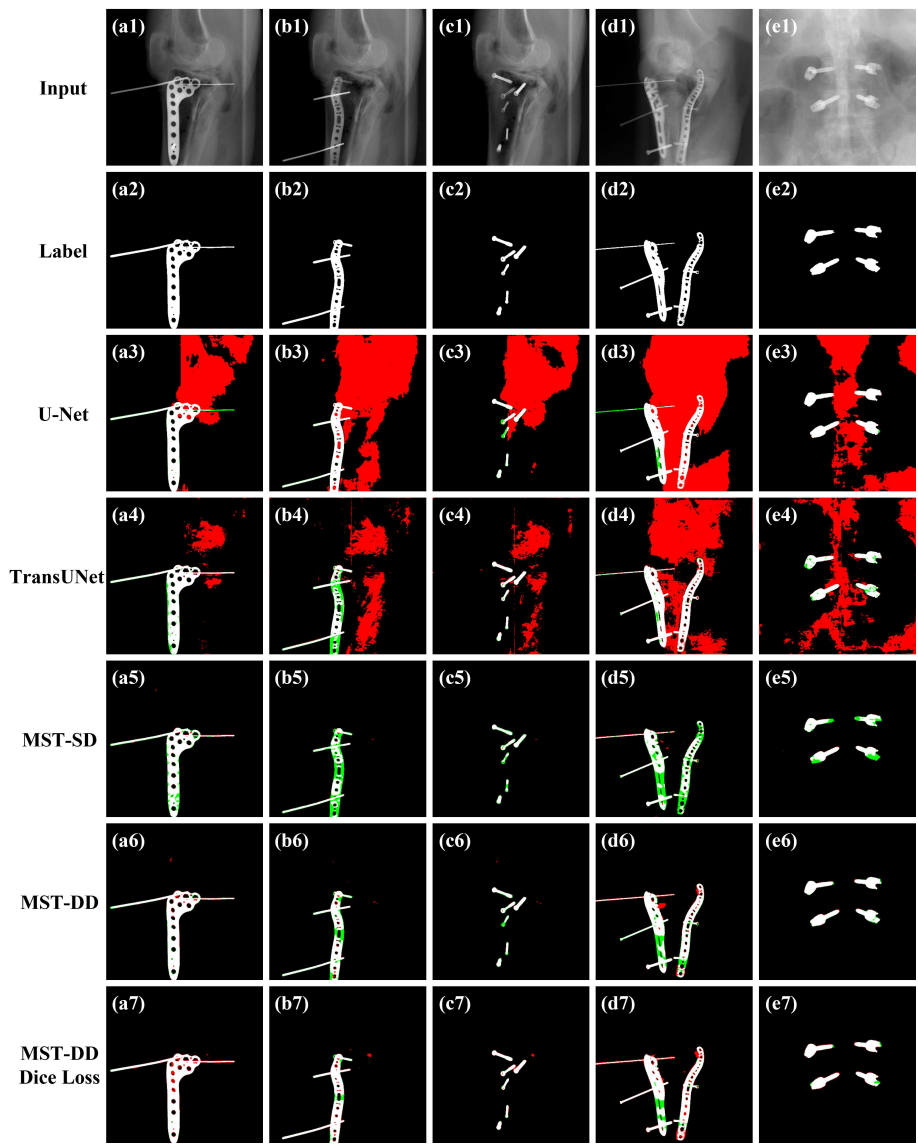}
\caption{Prediction results for real cadaver scans from different neural networks trained on simulated polychromatic projections. The true positive segmentation is labeled in white. The false positive and false negative segmentation are labeled in red and green, respectively.} \label{Results}
\end{figure}
The predictions for cadaver projections are displayed in Fig.~\ref{Results}. It shows four knee scans with different implants (Fig.~\ref{Results}(a1-d1)) and one spine scan with screws (Fig.~\ref{Results}(e1)). All models are trained on polychromatic projections. The intersected areas of predictions and labels are colored in white. The red and green areas stand for the false positives and false negatives, respectively. As it can be seen, The U-Net and TransUNet predict a large range of false positives near the bones. On the other hand, predictions of MST-SD have almost no false positives but still show incomplete segmentation of implants. By using the MST-DD and MST-DD with dice loss the performance is improved, with little false positives near the metal and less missing areas compared to MST-SD.

\section{Discussion}
Leveraging the high flexibility of metal simulation and insertion into real CT data enables us to generate large quantities of projection data. This can overcome the problem of data shortage for deep learning in the field of medical image processing. Another highly important aspect which supports the idea of simulated data is the high accuracy of segmentation labels.

The ViT started to draw attention since 2020. One important advantage is that it can transform between different tasks at the expense of requiring a large quantity of data for the initial training. After replacing the encoder from CNN to ViT, the performance of our setup greatly improved using the same training data, which can be testified by the results of MST-SD in Tab.~\ref{table_mono},\ref{table_poly} and Fig.~\ref{Results}. Further improvement is made after adding one more decoder, demonstrated by an  average dice score of MST-DD being above 0.85. The implementation of dice loss helps to improve the average recall in general and reduce the deviation for the mean dice score of MST-DD.

Evaluating the performance of networks trained on monochromatic and polychromatic projections, we can observe very limited improvement using polychromatic projections for U-Net and TransUNet. For the proposed MST-DD with dice loss, the segmentation performance is comparable for both data sets. ViT based encoder has less sensitivity and can focus on extracting the right features.

A limitation of this study is the reduction to the knee area as anatomical background. This could be generalized to the whole body in future work to make the network more robust. Another direction of research could be the fusion of simulated metal with real high resolution X-ray images instead of using synthetic images generated from CT volumes.


\section{Conclusion}
This work proposes the use of simulated training data combining CAD-based implants and CT volumes containing human anatomy. Aditionally we propose a dedicated network design MST-DD for metal segmentation. Combining the power of the large quantity of simulated images and the design of the ViT encoder and dual connected decoder enable us to construct a robust neural network which performs on real X-ray cadaver projections with an average dice score of 0.90.

\section{Disclaimer}
This work was supported by academic-industrial collaboration with Siemens Healthineers, XP Division. The presented method is not commercially available.

%
%

%
%
%
\bibliographystyle{splncs04}
%

\end{document}